# On the effect of force on DNA in the Peyrard-Bishop-Dauxois model

Likhachev I.V.[*], Lakhno V.D.

*The Institute of Mathematical Problems of Biology, Russian Academy of Sciences - a branch of the KIAM RAS. Address: IMPB RAS, 1, Professor Vitkevich St., 142290, Pushchino, Moscow Region, Russia*

**Abstract**

This paper presents a numerical study of the dynamics of DNA double helix breakage under the influence of external forces using the Peyrard–Bishop–Dauxois (PBD) model. The PBD model represents DNA as a chain of nonlinearly coupled oscillators, which makes it possible to analyze the processes of melting and mechanical denaturation. The main focus is on cases where an external force is applied to the terminal or central site of a DNA molecule, simulating stretching at a constant rate. The critical force required to break hydrogen bonds, which depends on the point of application of the force, is calculated. It is found that the denaturation process occurs stepwise, with characteristic peaks in the force-time graphs. The phenomenon of hysteresis under periodic exposure to external forces is also studied, which is important for understanding energy losses and heating of the system.

## 1. Introduction

DNA is one of the key molecules of life, providing storage and transmission of genetic information. However, its structure is not static: the DNA molecule is subjected to various mechanical and thermal influences, which can lead to its local or global denaturation. Studying the processes of hydrogen bond rupture in a double helix under the influence of external forces is important for understanding biological mechanisms such as replication and transcription, as well as for developing nanotechnology applications [1–9].

One of the approaches to modeling DNA dynamics is the Peyrard–Bishop–Dauxois (PBD) model [10,11], which simplifies the representation of DNA structure by considering a sequence of nucleotide pairs as a one-dimensional chain of nonlinearly coupled oscillators, where each base pair is modeled as a point mass interacting with neighboring pairs through a potential that takes into account hydrogen bonds and stacking interactions. This model makes it possible to effectively analyze the processes of melting and mechanical denaturation of DNA under various external conditions [12-14].

In this paper, we study the behavior of DNA at low temperatures under the influence of external forces aimed at transverse stretching of the chain. We plan to devote a separate article to consideration of final temperatures.

Our results advance the understanding of the DNA double helix's mechanical behavior, offering insights directly applicable to biophysical studies and to the growing fields of DNA nanotechnology and nanostructure-based devices. This work also uses viscous friction, but for this purpose, the Lemak–Balabaev thermostat [15–17] is used (see Appendix A).

The study is conducted at temperatures close to T = 0 K. Unlike the work [18], the state under consideration can be believed to be quasi-equilibrium with virtually no thermal motion due to the low cantilever velocity. We also note the works [19,20] devoted to dynamics and bubbles.

In this paper, direct one-dimensional modeling of molecular dynamics is used. Unlike [21], in which the potentials were approximated by a Taylor series, we use the original expressions for the potentials.

The paper is arranged as follows. Section 2 describes the model. Section 3 analyzes the data obtained in experiments on DNA stretching under a constant external force. Section 4 answers the question of whether the stretching process is reversible. Section 5 is devoted to conclusions.

Modeling the molecular dynamics of DNA double helix unwinding under the action of a directed external force is of fundamental importance for the interpretation and prediction of a wide range of biological and nanotechnological processes, including the work of DNA helicases during replication and transcription, mechanical chain separation in single-molecule optical and magnetic forceps experiments, as well as the process of controlled DNA sequence pulling and recognition in nanopore sequencing devices [22–25]. Numerical study of the dependence of the critical rupture force on the point of application and the rate of stretching [26] allows one not only to quantitatively describe the energy barriers associated with the rupture of hydrogen bonds and stacking interactions in various contexts [27], but also to identify such key features of the process as the stepwise nature of denaturation and hysteresis under cyclic loading [28], which is directly related to the thermodynamic efficiency of molecular machines [29] and the mechanical stability of DNA nanostructures developed for applications in nanomedicine and synthetic biology [30].

---

[*] Corresponding author, e-mail: ilya_lihachev@mail.ru.

## 2. Model and analytical methods

We consider the dynamics of DNA using the PBD model. In this model, after eliminating the center-of-mass motion, each nucleotide pair is represented by a single point mass.

Hamiltonian of the PBD model has the form [10,11]:

$$H = \sum_{n=1}^{N} \frac{M}{2} \dot{y}_n^2 + \sum_{n=1}^{N} U_{Morze}(y_n) + \sum_{n=1}^{N-1} U_{stack}(y_n, y_{n+1}), \quad (1)$$

where:

$$U_{Morze}(y_n) = D_n\left(1 - e^{-\alpha y_n}\right)^2, \quad (2)$$

$$U_{stack}(y_n) = \frac{k}{2}\left(1 + \rho e^{-\beta(y_n + y_{n+1})}\right)(y_n - y_{n+1})^2 \quad (3)$$

We choose the common effective mass as $M = 300.5$ a.m.u.; other parameters are taken from the paper by Campa and Giansanti: $\beta = 0.35$ Å$^{-1}$, $k = 0.025$ eV/Å$^2$, $\rho = 2.0$, $D_{AT} = 0.05$ eV, $\alpha_{AT} = 4.2$ Å$^{-1}$, $D_{GC} = 0.075$ eV, $\alpha_{GC} = 6.9$ Å$^{-1}$ [12].

In this model, M represents the effective mass of a nucleotide pair, $D_n$ and $\alpha$ are the depth and inverse width of a potential well describing the interaction of complementary nucleotides. The parameter k corresponds to an elastic constant that takes into account stacking interactions, β is the decay constant of these interactions, and ρ is a dimensionless coefficient that phenomenologically describes the cooperativity of denaturation in a π stack [7]. Summation is performed over all N nucleotide pairs, and in the PBD model, the value of $y_n$ reflects the relative displacement of the nucleotides of the n-th pair from their equilibrium position. In this model, $y_n$ corresponds to the actual elongation of hydrogen bonds in a complementary base pair multiplied by $\sqrt{2}$ (see [31]). The first term in (1) describes the kinetic energy of oscillations. The second term is the Morse potential $U_{Morze}$, which models the hydrogen bonds between nucleotides in a base pair. The third term corresponds to the stacking interaction of neighboring base pairs [10] $U_{stack}$.

If an external force is acting, the PBD Hamiltonian should involve the term:

$$H_F = -Fy, \quad (4)$$

where F is an external force causing a displacement y at any site in the chain.

## 3. Simulation results

To simulate the effect of a cantilever on DNA, the method of modeling molecular dynamics in its one-dimensional version was used. The potential was represented by equations (2) and (3) of intrasite and intersite interactions. An external force F=k·Δy (where k is the stiffness of the cantilever, and Δy the cantilever spring extension) was also added. The equations of motion were integrated using the Verlet speed algorithm. The temperature was maintained by the Lemak-Balabaev thermostat [15–17]. Also see appendix A.

The movement of the cantilever was always slower than the dynamic processes of breaking bonds in the chain.

Let us briefly describe the results of modeling the force action at low temperatures. There are various ways to apply a force to a DNA molecule. We will limit ourselves to considering two experimental cases in which an external force F is applied to the end sites of the DNA molecule (Fig. 1.2) and to the middle site (Fig. 1.1), where y is the deviation from the equilibrium between the nucleotide pairs at the point where force F is applied.

The external force F applied to the molecule arises from a spring connecting the terminal site to a certain point mass – the cantilever. The cantilever moves at a constant velocity. If the spring constant tends to infinity, then the terminal site to which the force is applied will also move at the same constant velocity.

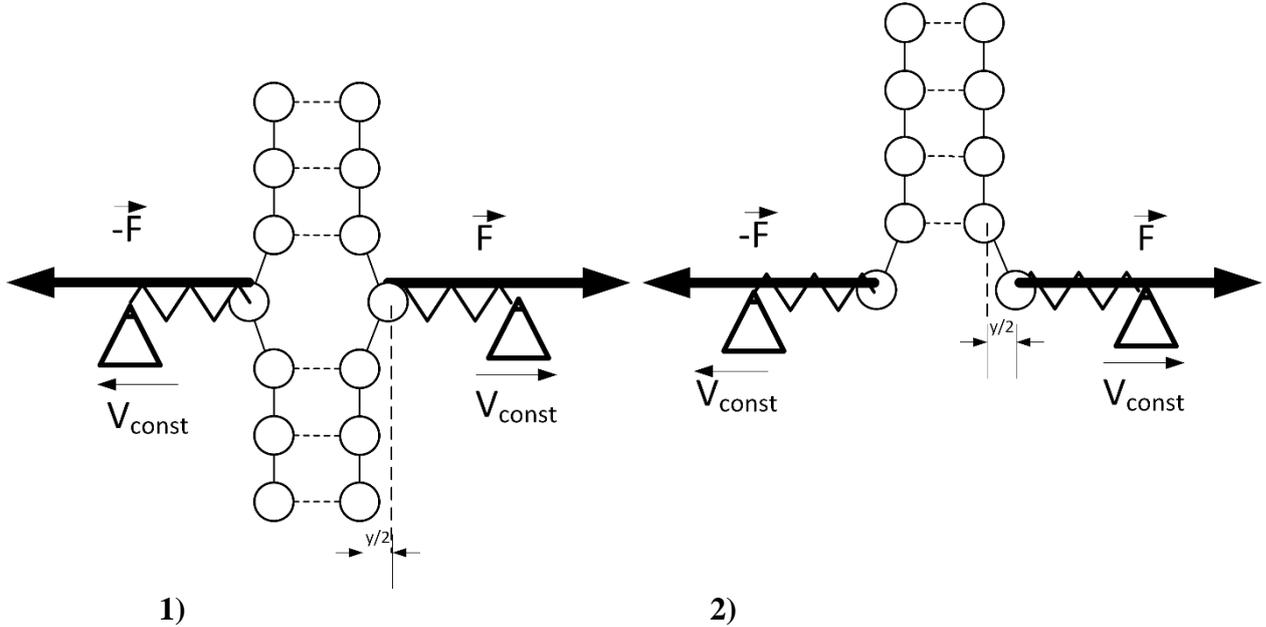

**Fig. 1.** Stretching of DNA for 1) the middle site with a force that ensures a constant stretching rate, 2) the initial site. Y is the coordinate of the site to which the external force is applied.

For the model under consideration (1), the force F consists of the sum of two forces: $F=F_{Morze}+F_{stack}$, where $F_{Morze}$ is the Morse force determined by the second term on the right-hand side of formula (1), $F_{stack}$ is the stacking force (Dauxois force) determined by the third term on the right-hand side of formula (1). If the force is applied to the middle site, the stacking force is approximately 2 times greater than the force applied to the terminal site, since the terminal site has only one adjacent nucleotide pair, while any non-terminal site has two (see Appendix B).

Figure 2 shows typical dependences of both the components $F_{Morze}$, $F_{stack}$, and the total force F acting on DNA on the displacement of y at the point of the force application.

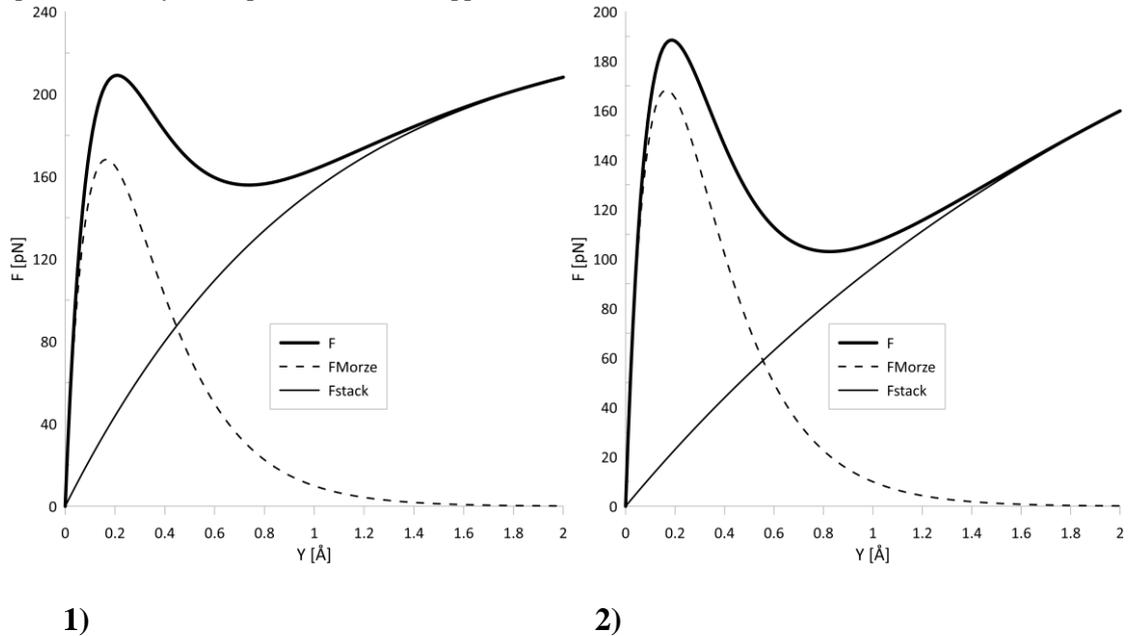

**Fig. 2.** Dependence of intra-site and inter-site interaction forces on the coordinate when stretching 1) the end site, 2) the middle site when fixing neighboring sites. DNA consists of 5 AT-pairs, $\beta = 0.35$ Å$^{-1}$, $k = 0.025$ eV/Å$^2$, $\rho = 2.0$, $D_{AT} = 0.05$ eV, $\alpha_{AT} = 4.2$ Å$^{-1}$. Y is the coordinate of the site to which the force is applied.

Here we present the dependence of force on displacement for a model with a fixed position of the rest of the chain. This allows us to make an approximate estimate for the critical force $F_c$, which leads to chain unzipping. Based on Graph 2, the force value at the maximum point for the operation parameters [12] is 187 pN if stretching occurs at the terminal site and 209 pN if stretching occurs at any middle site.

A typical dependence of the force on displacement y for a non-fixed chain is shown in Fig. 3, where F→0 as y→∞.

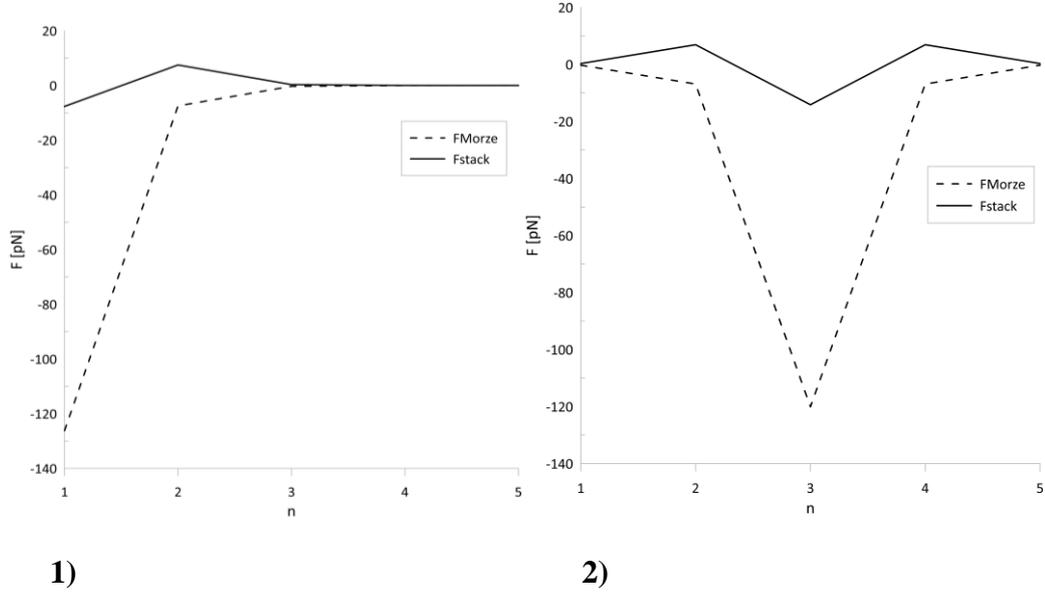

1)                                                                           2)

**Fig. 3.** Dependence of the force of intra-site interaction (dashes line) and inter-site interaction (solid line) on the site number when stretching a chain of 5 sites: 1) for the initial site, 2) for the third (middle) site. DNA consists of 5 AT-pairs, $\beta = 0.35$ Å$^{-1}$, $k = 0.025$ eV/Å$^2$, $\rho = 2.0$, $D_{AT} = 0.05$ eV, $\alpha_{AT} = 4.2$ Å$^{-1}$.

Figure 3 shows the distribution of the force across the sites of a five-base-pair chain, for $F<F_c$ for the values of the chain parameters from [12]. $F_c$ is the critical force at which the chain unzips, it coincides with the maximum force exerted by the cantilever. For the first site (if this site is terminal: left), this force is equal in modulus and opposite in sign to the sum of the forces $F_{Morze1}$ (the Morse potential force) and $F_{stack12}$ (the force of inter-site interaction acting from the second site):

$$F_{c,left} = \frac{\partial U_{Morze}(y_1)}{\partial y_1} + \frac{\partial U_{stack}(y_1, y_2)}{\partial y_1} \quad (5)$$

In turn, the total force $F_2$ acting on the second site is equal to $F_2=F_{Morze2}+F_{stack21}$, where $F_{Morze2}$ is the force of the Morse potential, $F_{stack21}$ is the force of the inter-site potential acting from the first and third sites.

For any site with number m from the internal part of the chain, the total force will be:

$$F_{c,middle} = \frac{\partial U_{Morze}(y_m)}{\partial y_m} + \frac{\partial U_{stack}(y_{m-1}, y_m)}{\partial y_m} + \frac{\partial U_{stack}(y_m, y_{m+1})}{\partial y_m}, \quad 1<m<N \quad (6)$$

The choice of the cantilever spring constant does not affect the maxima of the $F_c$ forces obtained at $T = 0$ K. This is not the case at temperatures other than 0 K. Appendix B shows that the choice of the spring constant affects only the DNA dynamics after passing each peak of the force.

Figures 4 and 5 show the results of modeling the unwinding of a chain in two cases: when a force is applied to the terminal site of the chain (Fig.4) and to the central site (Fig.5). The sawtooth curves in Figures 4 and 5 correspond to the sequential breaking of stacked bonds with increasing y (and correspondingly with increasing time). The nearly vertical lines in Figure 4 show that the bond breaking processes are very fast (on the order of several picoseconds; see Appendix C).

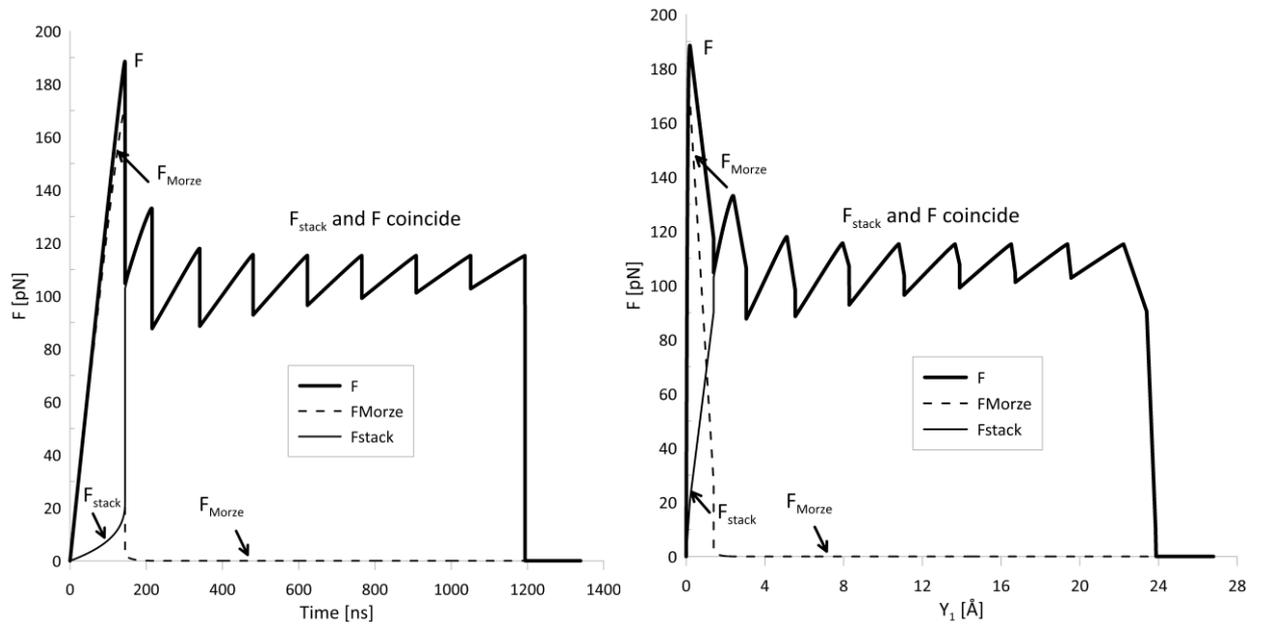

**Fig. 4.** Stretching of DNA by the first terminal site with a force that ensures a constant cantilever velocity. The total force F and its components are: $F_{Morse}$ is the force of the Morse potential, $F_{stack}$ is the force of inter-site interaction. On the left is the dependence of force (F) on time, on the right is the dependence of force (F) on the coordinate of the first site ($Y_1$) (the coordinates of the site where stretching occurs). DNA consists of 10 AT-pairs, $\beta = 0.35$ Å$^{-1}$, $k = 0.025$ eV/Å$^2$, $\rho = 2.0$, $D_{AT} = 0.05$ eV, $\alpha_{AT} = 4.2$ Å$^{-1}$.

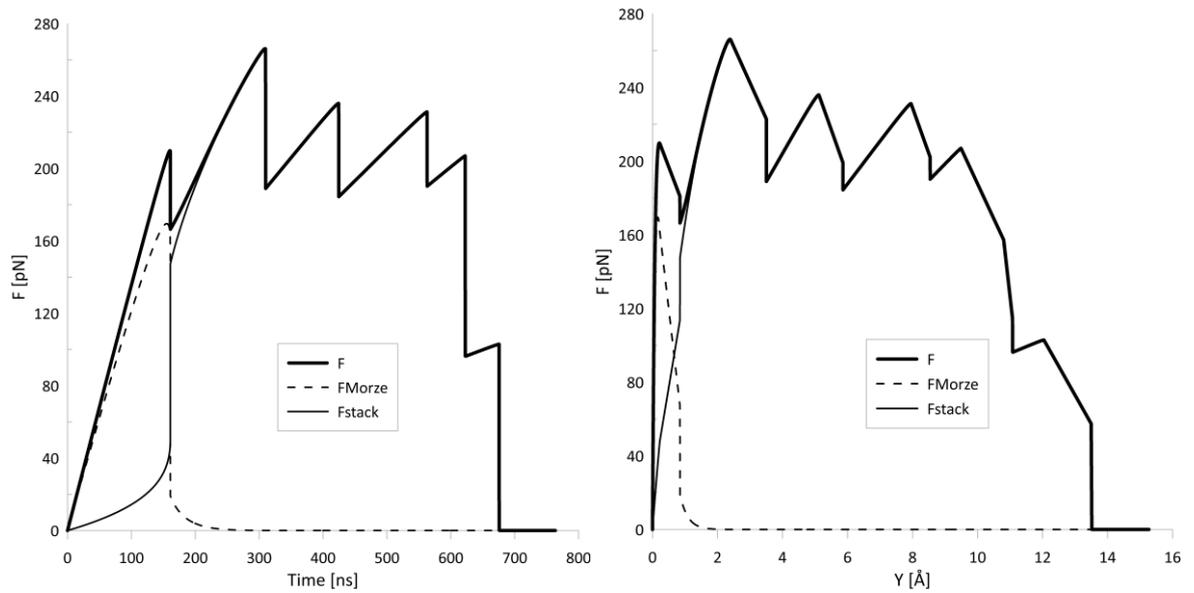

**Fig. 5.** Stretching of DNA by the central site with a force that ensures a constant cantilever velocity. The total force F and its components are: $F_{Morse}$ is the force of the Morse potential, $F_{stack}$ is the force of inter-site interaction. On the left is the dependence of force (F) on time, on the right is the dependence of force (F) on the coordinate of a middle site (Y) (the coordinates of the site where stretching occurs). DNA consists of 10 AT-pairs, $\beta = 0.35$ Å$^{-1}$, $k = 0.025$ eV/Å$^2$, $\rho = 2.0$, $D_{AT} = 0.05$ eV, $\alpha_{AT} = 4.2$ Å$^{-1}$.

According to [26], the value of $F_c$ when the point of application of the force is located far from the ends of the chain is equal to 239.43 pN, and when the force is applied to the terminal site, $F_c$ will be 226.48 pN. The difference in value is 12.95 pN. The calculations in this paper show that both the terminal and the middle site are affected by one intra-site Morse potential force (168.3 pN) and one (in the case of the terminal site) inter-site potential force (18.7 or 19.5 pN, depending on the end of DNA 3` or 5`), or two inter-site potential forces (18.7 + 19.5 = 38.2 pN). Thus, for the terminal site, a force of 187 pN is obtained in the case of a 3' end and 187.8 pN in the case of a 5' end. If one pulls on the middle site, the force will be 206.5 pN.

## 4. Hysteresis

In this section, we will consider the question: what happens if a cantilever is applied to a system periodically? Will the force characteristic of the DNA on the return path be exactly the same as that during stretching?

Let us consider the quasi-static periodic motion of a cantilever. We will choose 1 μs as the period. Let us assume that the cantilever moves half a period in the direction of DNA stretching and half a period in the direction of the initial state so that the chain does not unzip. Thus, after one full period (or n full periods, where n is a natural number), the cantilever will be at the coordinate $Y_{cant} = 0$ Å.

Then the dependence of the force on time and on the coordinate will look like this:

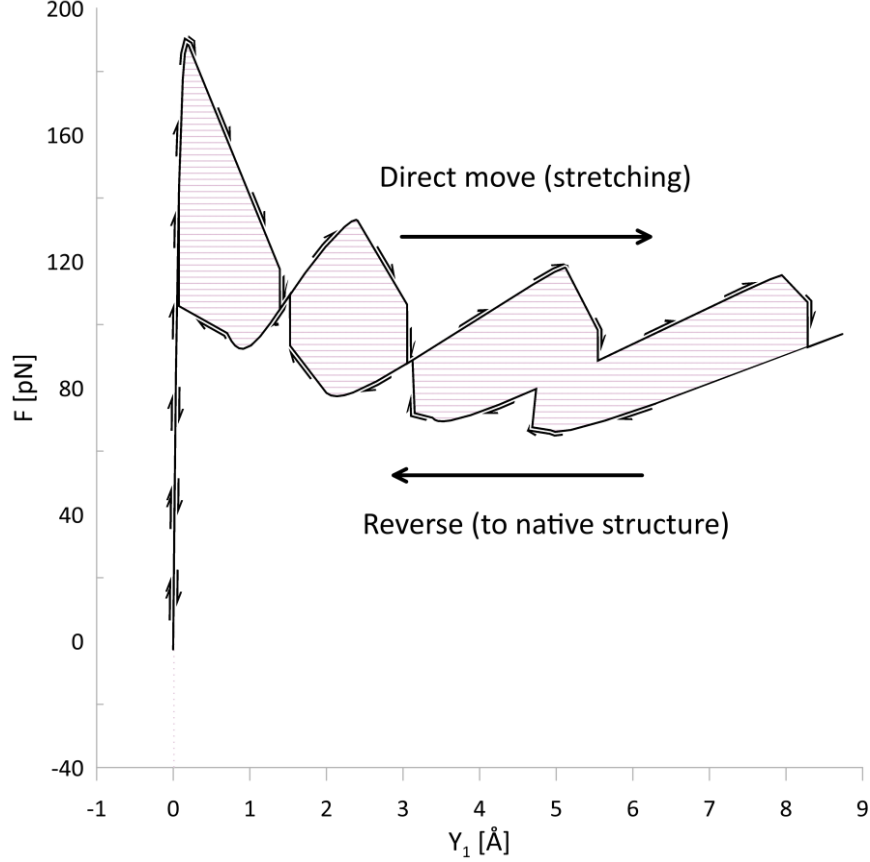

**Fig. 6.** The phenomenon of hysteresis under the action of a periodic force from the cantilever: the dependence of the force on the coordinate of the first site.

In this case, the integral of the force along the hysteresis contour determines the energy loss of the DNA chain Q during one hysteresis cycle.:

$$Q = \oint F(y)dy \qquad (7)$$

To evaluate integral (7), a cyclic experiment is performed: for 1 μs, the chain is stretched by 10 Å, breaking the hydrogen bonds between the base pairs, and then compressed for another 1 μs. In this case, hydrogen bonds arise again. The force exerted by the cantilever during stretching of the molecule is greater than that during compression. "Compression" refers to the reverse movement of the cantilever, not to the force compression.

Hysteresis losses lead to heating of the chain, which must be taken into account when creating structures for generating mechanical oscillations in DNA.

If $Q_1$ is the energy expended by the cantilever when stretching DNA, and $Q_2$ is the energy obtained by the cantilever during its return stroke, then $Q=Q_1-Q_2$. Accordingly, for the parameters we use, the efficiency coefficient will be: $\eta = \dfrac{Q_1 - Q_2}{Q_1}$, $\eta = 0.2332$. Note that the efficiency calculation is given for a specific case. If a different stretching-compression period had been chosen for the same sample, the efficiency could have been different.

## 5. Conclusions

In this paper, a numerical study of the dynamics of DNA double helix breakage under the influence of external forces at low temperatures was carried out within the framework of the Peyard–Bishop–Dauxois model for the parameter values of [12]. The case of physiological temperatures for the parameters of [12] in the thermodynamic equilibrium case was considered in [13]. We considered stretching at a constant rate. The analysis of the results allowed us to draw a number of key conclusions.

1. It was revealed that the breakage of the DNA double helix occurs in stages, with the first breaks requiring significantly greater forces than the subsequent ones. This is due to the peculiarities of the distribution of forces between the nucleotide pairs.
2. It was found that, at Campa parameters, the critical force required to initiate a rupture is approximately 210 pN, which is consistent with previously published data. However, a further increase in the applied force to 266 pN leads to an avalanche-like denaturation process.
3. When a force is applied that ensures a constant rate of stretching, characteristic peaks are observed on the graph of force versus time, which indicates a sequential rupture of individual sections of the chain.
4. It was found that the mechanisms of denaturation vary depending on the external influence applied: at a constant force, a sudden elongation of the chain is observed, whereas at a constant velocity the process proceeds more uniformly.
5. Slow movement of the cantilever leads to rapid dynamic changes in the chain.
6. Microsecond periodic movement of the cantilever leads to picosecond dynamics in the chain.

Numerous examples of a force acting on DNA can be found in living cells, where double-stranded DNA and folded RNA are moved and unzipped by enzymes [32]. Another example is DNA unzipping in nanopores, which is important in genome sequencing setups in the processes of capturing and pulling DNA through nanopores [23–25,33]. Thus, experiments on unzipping in combination with theoretical methods have made it possible to describe the abrupt unzipping transition and the thermodynamics of RNA motifs or DNA pins [34,35]. Our molecular dynamic modeling can form the basis for describing such processes.

In particular, we demonstrated that cantilever oscillations in the terahertz frequency range will be accompanied by significant thermal losses. The efficiency of a specific DNA stretching-compression cycle using a cantilever was calculated.

The results of the work may be useful for further study of the mechanical stability of DNA, as well as for the development of methods for controlled denaturation in biotechnological and nanotechnological applications.

## 6. Acknowledgments

The authors thank N.K. Balabaev for valuable comments during the discussion of the work.
The calculations were performed on the hybrid supercomputer K60 installed in the Supercomputer Centre of Collective Usage of KIAM RAS.
This work was supported by Keldysh Institute of Applied Mathematics RAS [FFMN-2025-0024].

**References**

[1] J. Marmur, P. Doty, Determination of the base composition of deoxyribonucleic acid from its thermal denaturation temperature, Journal of Molecular Biology 5 (1962) 109–118. https://doi.org/10.1016/S0022-2836(62)80066-7.
[2] M.D. Wang, H. Yin, R. Landick, J. Gelles, S.M. Block, Stretching DNA with optical tweezers, Biophysical Journal 72 (1997) 1335–1346. https://doi.org/10.1016/S0006-3495(97)78780-0.
[3] C. Bustamante, S.B. Smith, J. Liphardt, D. Smith, Single-molecule studies of DNA mechanics, Current Opinion in Structural Biology 10 (2000) 279–285. https://doi.org/10.1016/S0959-440X(00)00085-3.
[4] P.H. von Hippel, E. Delagoutte, A General Model for Nucleic Acid Helicases and Their "Coupling" within Macromolecular Machines, Cell 104 (2001) 177–190. https://doi.org/10.1016/S0092-8674(01)00203-3.
[5] N.C. Seeman, DNA in a material world, Nature 421 (2003) 427–431. https://doi.org/10.1038/nature01406.
[6] A.V. Pinheiro, D. Han, W.M. Shih, H. Yan, Challenges and opportunities for structural DNA nanotechnology, Nature Nanotech 6 (2011) 763–772. https://doi.org/10.1038/nnano.2011.187.
[7] D. Salerno, A. Tempestini, I. Mai, D. Brogioli, R. Ziano, V. Cassina, F. Mantegazza, Single-Molecule Study of the DNA Denaturation Phase Transition in the Force-Torsion Space, Phys. Rev. Lett. 109 (2012) 118303. https://doi.org/10.1103/PhysRevLett.109.118303.
[8] A.E. Bergues-Pupo, J.M. Bergues, F. Falo, Unzipping of DNA under the influence of external fields, Physica A Statistical Mechanics and Its Applications 396 (2014) 99–107. https://doi.org/10.1016/j.physa.2013.10.050.
[9] A.E. Bergues-Pupo, J.M. Bergues, F. Falo, A. Fiasconaro, Thermal and inertial resonances in DNA unzipping, Eur. Phys. J. E 38 (2015) 41. https://doi.org/10.1140/epje/i2015-15041-4.

[10] T. Dauxois, M. Peyrard, A.R. Bishop, Entropy-driven DNA denaturation, Phys. Rev. E 47 (1993) R44–R47. https://doi.org/10.1103/PhysRevE.47.R44.

[11] T. Dauxois, M. Peyrard, A.R. Bishop, Dynamics and thermodynamics of a nonlinear model for DNA denaturation, Phys. Rev. E 47 (1993) 684–695. https://doi.org/10.1103/PhysRevE.47.684.

[12] A. Campa, A. Giansanti, Experimental tests of the Peyrard-Bishop model applied to the melting of very short DNA chains, Phys. Rev. E 58 (1998) 3585–3588. https://doi.org/10.1103/PhysRevE.58.3585.

[13] N.K. Voulgarakis, A. Redondo, A.R. Bishop, K.Ø. Rasmussen, Probing the Mechanical Unzipping of DNA, Phys. Rev. Lett. 96 (2006) 248101. https://doi.org/10.1103/PhysRevLett.96.248101.

[14] A. Sulaiman, F.P. Zen, H. Alatas, L.T. Handoko, Dynamics of DNA breathing in the Peyrard–Bishop model with damping and external force, Physica D: Nonlinear Phenomena 241 (2012) 1640–1647. https://doi.org/10.1016/j.physd.2012.06.011.

[15] N.K. Balabaev, A.S. Lemak, Molecular dynamics of a linear polymer in a hydrodynamic flow, Russian Journal of Physical Chemistry A 69 (1995). https://elibrary.ru/item.asp?id=21474281 (accessed August 22, 2020).

[16] A.S. Lemak, N.K. Balabaev, A comparison between collisional dynamics and brownian dynamics, Molecular Simulation 15 (1995). https://doi.org/10.1080/08927029508022336.

[17] A.S. Lemak, N.K. Balabaev, Molecular dynamics simulation of a polymer chain in solution by collisional dynamics method, Journal of Computational Chemistry 17 (1996). https://doi.org/10.1002/(SICI)1096-987X(19961130)17:15%253C1685::AID-JCC1%253E3.0.CO;2-L.

[18] M. Hillebrand, G. Kalosakas, Ch. Skokos, A.R. Bishop, Distributions of bubble lifetimes and bubble lengths in DNA, Phys. Rev. E 102 (2020) 062114. https://doi.org/10.1103/PhysRevE.102.062114.

[19] DL_POLY: Application to molecular simulation: Molecular Simulation: Vol 28, No 5, (n.d.). https://www.tandfonline.com/doi/abs/10.1080/08927020290018769 (accessed May 6, 2025).

[20] M. Hillebrand, G. Kalosakas, A.R. Bishop, Ch. Skokos, Bubble lifetimes in DNA gene promoters and their mutations affecting transcription, The Journal of Chemical Physics 155 (2021) 095101. https://doi.org/10.1063/5.0060335.

[21] A. Djine, N.O. Nfor, G.R. Deffo, S.B. Yamgoué, Higher order investigation on modulated waves in the Peyrard–Bishop–Dauxois DNA model, Chaos, Solitons & Fractals 181 (2024) 114706. https://doi.org/10.1016/j.chaos.2024.114706.

[22] M. Jain, S. Koren, K.H. Miga, J. Quick, A.C. Rand, T.A. Sasani, J.R. Tyson, A.D. Beggs, A.T. Dilthey, I.T. Fiddes, S. Malla, H. Marriott, T. Nieto, J. O'Grady, H.E. Olsen, B.S. Pedersen, A. Rhie, H. Richardson, A.R. Quinlan, T.P. Snutch, L. Tee, B. Paten, A.M. Phillippy, J.T. Simpson, N.J. Loman, M. Loose, Nanopore sequencing and assembly of a human genome with ultra-long reads, Nat Biotechnol 36 (2018) 338–345. https://doi.org/10.1038/nbt.4060.

[23] E.A. Manrao, I.M. Derrington, A.H. Laszlo, K.W. Langford, M.K. Hopper, N. Gillgren, M. Pavlenok, M. Niederweis, J.H. Gundlach, Reading DNA at single-nucleotide resolution with a mutant MspA nanopore and phi29 DNA polymerase, Nat Biotechnol 30 (2012) 349–353. https://doi.org/10.1038/nbt.2171.

[24] A.F. Sauer-Budge, J.A. Nyamwanda, D.K. Lubensky, D. Branton, Unzipping Kinetics of Double-Stranded DNA in a Nanopore, Phys. Rev. Lett. 90 (2003) 238101. https://doi.org/10.1103/PhysRevLett.90.238101.

[25] J. Comer, V. Dimitrov, Q. Zhao, G. Timp, A. Aksimentiev, Microscopic mechanics of hairpin DNA translocation through synthetic nanopores, Biophys J 96 (2009) 593–608. https://doi.org/10.1016/j.bpj.2008.09.023.

[26] N. Singh, Y. Singh, Statistical theory of force-induced unzipping of DNA, Eur. Phys. J. E 17 (2005) 7–19. https://doi.org/10.1140/epje/i2004-10100-7.

[27] C. Bustamante, Y.R. Chemla, N.R. Forde, D. Izhaky, Mechanical processes in biochemistry, Annu Rev Biochem 73 (2004) 705–748. https://doi.org/10.1146/annurev.biochem.72.121801.161542.

[28] J. Liphardt, B. Onoa, S.B. Smith, I. Tinoco, C. Bustamante, Reversible unfolding of single RNA molecules by mechanical force, Science 292 (2001) 733–737. https://doi.org/10.1126/science.1058498.

[29] Fluctuation driven ratchets: Molecular motors | Phys. Rev. Lett., (n.d.). https://journals.aps.org/prl/abstract/10.1103/PhysRevLett.72.1766 (accessed November 25, 2025).

[30] F. Hong, F. Zhang, Y. Liu, H. Yan, DNA Origami: Scaffolds for Creating Higher Order Structures, Chem Rev 117 (2017) 12584–12640. https://doi.org/10.1021/acs.chemrev.6b00825.

[31] M. Santosh, P.K. Maiti, Force induced DNA melting, J Phys Condens Matter 21 (2009) 034113. https://doi.org/10.1088/0953-8984/21/3/034113.

[32] A.M. Pyle, Translocation and unwinding mechanisms of RNA and DNA helicases, Annu Rev Biophys 37 (2008) 317–336. https://doi.org/10.1146/annurev.biophys.37.032807.125908.

[33] Nanopore sequencing and assembly of a human genome with ultra-long reads - PubMed, (n.d.). https://pubmed.ncbi.nlm.nih.gov/29431738/ (accessed March 12, 2025).

[34] Experimental validation of free-energy-landscape reconstruction from non-equilibrium single-molecule force spectroscopy measurements | Nature Physics, (n.d.). https://www.nature.com/articles/nphys2022 (accessed March 12, 2025).

[35] O.K. Dudko, J. Mathé, A. Szabo, A. Meller, G. Hummer, Extracting Kinetics from Single-Molecule Force Spectroscopy: Nanopore Unzipping of DNA Hairpins, Biophysical Journal 92 (2007) 4188–4195. https://doi.org/10.1529/biophysj.106.102855.

# 1. Appendix A

Lemak–Balabaev collision thermostat

For thermostatting systems in molecular dynamics, especially at high temperatures or under conditions of weak coupling with a thermostat, the **Lemak–Balabaev collisional thermostat** is effectively used [1,2]. A review of the methods of temperature maintenance in canonical ensembles is given in [3]. In this method, thermal equilibrium is achieved due to periodic collisions of particles of the system with fictitious thermostat particles, whose velocities are randomly selected from the Maxwellian distribution at a given temperature.

Each particle of the system with a certain probability collides with a virtual particle of the thermostat, as a result of which its velocity changes according to the law of elastic collisions. These collisions occur as a stochastic process, which avoids artifacts inherent in deterministic thermostats, such as distortions in velocity or energy distributions.

The advantages of the Lemak-Balabaev thermostat are:
- correct observance of the Maxwell-Boltzmann distribution law;
- stability in modeling nonequilibrium processes;
- possibility of local temperature control in the selected subsystem.

This thermostat is especially useful when modeling systems with volatile components (for example, when studying desorption, sublimation, evaporation, etc.), where it is necessary to maintain realistic particle behavior at the boundary between heated and cold regions.

**Main parameters**
- $\nu$ — collision frequency (stochastic intensity, units $s^{-1}$);
- $\Delta t$ — integration time step;
- T — preset thermostat temperature;
- $m_i$ — mass of particle i;
- $v_i$ — velocity of particle i.

**Collision probability**

For each particle at each time step, a check is made to determine whether a collision with the thermostat will occur. The collision probability is given by the formula:
$$P = 1 - exp(-\nu \Delta t)$$
If the random number r∈[0,1], generated in step satisfies r<P, then a collision is considered to have occurred.

**Collision rule**

As a result of a collision, the velocity of the particle changes according to the law of elastic collision with a thermostat particle, the velocity of which is taken from the Maxwell distribution:
$$v'_i = v_i - \frac{2m_T}{m_i + m_T}(v_i - v_T)$$
where:
- $m_T$ – mass of a fictitious thermostat particle;
- $v_T$ – the velocity of a thermostat particle is chosen randomly from the Maxwell distribution at temperature T:

$$P(v_T) \propto \exp\left(-\frac{m_T v_T^2}{2k_B T}\right)$$

That is after a collision, the particle's velocity is simply replaced by a new one chosen from the Maxwellian distribution.

The zero initial temperature in the system is achieved by setting zero velocities for all sites. The velocity of the cantilever is so low that the kinetic energy of the system is close to zero. After passing the peak of the force, the nucleotide can accelerate towards a local minimum of potential energy, heating the system, as well as transferring its kinetic energy to neighboring nucleotides, generating an oscillation wave. This motion is dampened due to the use of a collision thermostat with sufficient viscosity [1,2,4]. The oscillation attenuation time is several picoseconds, depending on the viscosity of the collision thermostat (see Appendix C). The particles of the virtual collision medium, having zero velocity, absorb the kinetic energy of the nucleotide, thereby providing a temperature close to 0 K.

# 2. Appendix B

The energy determined by equation (1) has two minima, the lower of which is the one corresponding to smaller values of y. As y increases, the lower-energy minimum becomes the second minimum. Equilibrium in this case becomes impossible, and the nucleotide pairs, now held together only by the stacking force, begin to open. A peculiar wave of chain unzipping occurs, in which a jump in force occurs each time the next nucleotide pair opens. In this case, the dependence of the absolute value of the force on the displacement y describes a barrier for small y, and far from the barrier in an infinite chain, a steady-state force value is set F(y=∞)=0 if the chain is unzipped, or F<Fc, if the force value is less than the critical value [5].

To the stiffness of the cantilever spring. It only affects the DNA dynamics after each peak of strength has passed. The rupture of the Morse potential (under the influence of the stacking interaction) occurs at a specific, well-defined external force. This critical rupture force ($F_c$) is measured as the product of the cantilever's spring constant (k) and its deflection at the moment of rupture (Δx), i.e., $F_c = k \cdot \Delta x$. Only the value of $F_c$ is critical; the product k·Δx merely affects the time (or cantilever position) at which rupture occurs.

The initial temperature in the system is zero due to setting zero velocities for all sites. The velocity of the cantilever is so low that the kinetic energy of the system is close to zero. After passing the peak of force, the nucleotide can accelerate toward a local minimum of potential energy, heating the system and transferring its kinetic energy to neighboring nucleotides, generating an oscillation wave. This motion is dampened due to the use of a collision thermostat with sufficient viscosity[1,2,4]. The particles in the virtual collision medium, having zero velocity, absorb the kinetic energy of the nucleotide, thereby maintaining a temperature close to 0 K.

**Calculation the force in the peak point of Morze potential in the case of fixation of neighboring DNA sites**

The maximum force exerted by the Morse potential is:

$$U_{Morze} = D\left(e^{-ay}-1\right)^2$$

$$F_{Morze} = -\frac{dU_m}{dy} = 2Da \cdot e^{-ay} \cdot (e^{-ay} - 1)$$

Here, $U_{Morse}$ and $F_{Morse}$ denote the on-site Morse potential and the corresponding force.

The on-site interaction is modeled by the Morse potential. Its minimum is found by taking the first derivative to obtain the force (with a sign reversal) and then finding the critical point of this force function by setting its derivative to zero. The solution yields the coordinate of the potential minimum.

$$\frac{dF}{dy} = -2D \cdot a \cdot e^{-2ar} - 2D \cdot a \cdot e^{-a*r} \cdot (e^{-ar} - 1)$$

$$y_{max} = -\frac{\ln(0.5)}{a} = 0.165 \text{ Å}$$

$$F_m(y_{max}) = 168.3 \text{ pN}$$

Assume that the displacement of the next site is zero, then:

$$U_{stack} = k * 0.5 * (1 + \rho \cdot e^{-b*(y_1+y_2)}) \cdot (y_1-y_2)^2$$

Differentiation of the stacking potential can be performed either over coordinate $y_1$ or $y_2$ of the two neighboring sites. It is a notable (and perhaps counterintuitive) aspect of the PBD model that the forces derived from this potential act differently on the first and the second site.

$$F_{stack1} = -\frac{dW}{dy_1} = -k \cdot (y_1-y_2) \cdot (1. + \rho \cdot e^{-b(y_1+y_2)}) + b \cdot k \cdot 0.5 \cdot \rho \cdot e^{-b(y_1+y_2)} \cdot (y_1-y_2)^2 = 18,7 \, pN$$

$$F_{stack2} = -\frac{dW}{dy_2} = +k \cdot (y_1-y_2) \cdot (1. + \rho \cdot e^{-b(y_1+y_2)}) + b \cdot k \cdot 0.5 \cdot \rho \cdot e^{-b(y_1+y_2)} \cdot (y_1-y_2)^2 = 19.5 \, pN$$

$$F_{all} = F + 2F_{stack} = 168.3 \, pN + 18.7 \, pN + 19.5 \, pN = 206,5 \, pN$$

Here, $U_{stack}$ and $F_{stack}$ represent the potential and force of the inter-base-pair (stacking) interaction.

This result is consistent with Figure 2 of [5], both in terms of position and magnitude of force.

According to the inter-site interaction potential, the forces acting on two neighboring sites are not equal. Let us look at formulas $F_{stack1}$ and $F_{stack2}$. For the forces of inter-site interaction, the second term is the same, but the first term has a different sign. Because of this, these forces are not equal in the PBD model.

## 3. Appendix C

After passing the force peak and continuing to operate the cantilever, the hydrogen bonds between nucleotide pairs are broken. This increases the inter-site interaction forces of the next nucleotide pair. Due to the collision thermostat, the system experiences damped oscillations. The higher the viscosity of the medium, the faster the oscillations decay. Figure C.1 shows two types of damped oscillations: fast – at high viscosity of the virtual collision medium (at λ=100) and slow – at low viscosity (at λ=0.01).

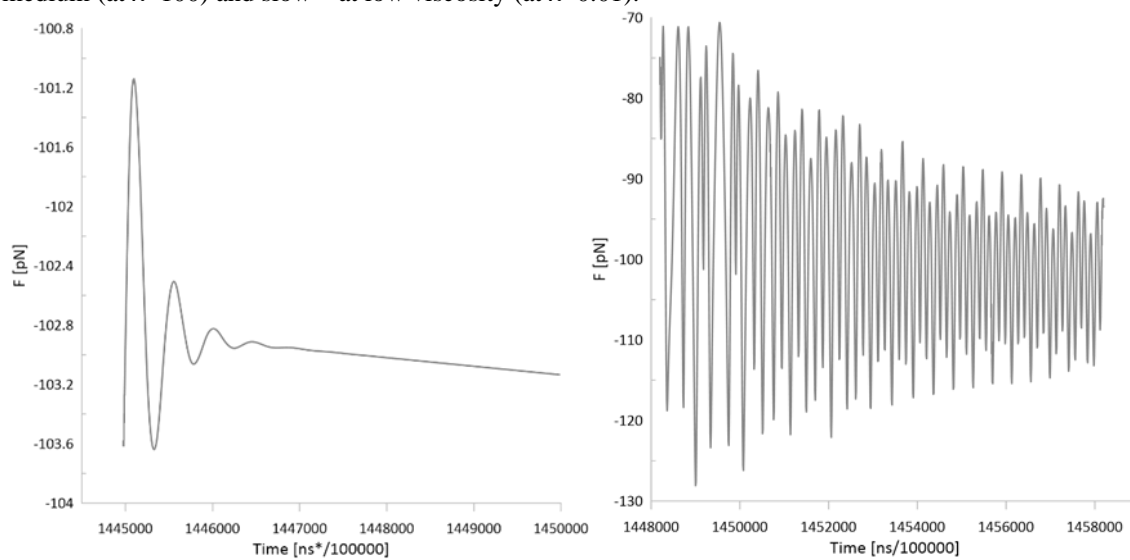

**Figure C.1.** The dependence of the force on time after the rupture of hydrogen bonds at high viscosity of the virtual collision medium (left, at λ=100) and at low viscosity (right, at λ=0.01)

## References

[1] A.S. Lemak, N.K. Balabaev, A comparison between collisional dynamics and brownian dynamics, Molecular Simulation 15 (1995). https://doi.org/10.1080/08927029508022336.

[2] A.S. Lemak, N.K. Balabaev, Molecular dynamics simulation of a polymer chain in solution by collisional dynamics method, Journal of Computational Chemistry 17 (1996). https://doi.org/10.1002/(SICI)1096-987X(19961130)17:15%253C1685::AID-JCC1%253E3.0.CO;2-L.

[3] G.A. Vinogradov, V.D. Lakhno, On the Thermalization of One-Dimensional lattices. II. Canonical Ensemble, Math.Biol.Bioinf. 20 (2025) 31–46. https://doi.org/10.17537/2025.20.31.

[4] N.K. Balabaev, A.S. Lemak, Molecular dynamics of a linear polymer in a hydrodynamic flow, Russian Journal of Physical Chemistry A 69 (1995). https://elibrary.ru/item.asp?id=21474281 (accessed August 22, 2020).

[5] A. Singh, B. Mittal, N. Singh, Force induced unzipping of dsDNA: The solvent effect, Phys. Express 3 (2013) 18. https://www.researchgate.net/profile/Purna-Baral/publication/313900210_Spin_susceptibility_A_study_of_anomalies_due_to_Kondo_effect_and_f-electron_correlation_in_HF_systems/links/58aef3d0aca2725b54111959/Spin-susceptibility-A-study-of-anomalies-due-to-Kondo-effect-and-f-electron-correlation-in-HF-systems.pdf#page=23 (accessed April 1, 2025).